\documentclass[aps,pra,twocolumn,groupedaddress]{revtex4-1}
	\usepackage{amsmath,bm}
	\usepackage{amssymb}
	\usepackage[colorlinks,bookmarks=false,citecolor=blue,linkcolor=black,urlcolor=blue]{hyperref}

\newcommand{\abs}[1]{\left|{#1}\right|}
\newcommand{\ket}[1]{|{#1}\rangle}
\newcommand{\bra}[1]{\langle{#1}|}
\newcommand{\braket}[2]{\langle{#1}|{#2}\rangle}
\newcommand{\tr}{\text{Tr}}

\newcommand{\Tr}{\text{Tr}}

\newcommand{\rinitial}{\Psi}
\newcommand{\expectT}{\overline{\text{T}}_{\rinitial}}
\newcommand{\limes}[1]{\lim\limits_{#1 \to \infty}}

\newcommand{\SSS}[1]{\mathcal{S}\left[ #1\right]}
\newcommand{\sss}{\mathcal{S}}
\newcommand{\ddd}{\mathcal{D}}
\newcommand{\ttt}{\mathcal{T}_{\perp}}
\newcommand{\SSSn}[1]{\mathcal{S}^t\left[ #1\right]}
\newcommand{\MS}[1]{\mathcal{MS}\left[ #1\right]}
\newcommand{\MSt}[1]{(\mathcal{MS})^t\left[ #1\right]}
\newcommand{\MMM}[1]{\mathcal{M}\left[ #1\right]}
\newcommand{\mmm}{\mathcal{M}}

\newcommand{\hrho}{\hat{\varrho}}
\newcommand{\hK}{\hat{K}}
\newcommand{\hA}{\hat{A}}
\newcommand{\hI}{\hat{\mathbb{I}}}
\newcommand{\hchi}{\hat{\chi}}
\newcommand{\rcond}{\hrho_{\text{cond}}}
\newcommand{\hsigma}{\hat{\sigma}}
\newcommand{\phiD}{\varphi_\mathcal{D}}
\newcommand{\phiR}{\varphi_\mathcal{R}}
\newcommand{\PsiD}{\Psi_\mathcal{D}}
\newcommand{\PsiR}{\Psi_\mathcal{R}}
\newcommand{\cD}{c_\mathcal{D}}
\newcommand{\cR}{c_\mathcal{R}}
\newcommand{\chim}{\hat{\chi}_\mmm}

\begin{document}
\title{Generalized Kac's Lemma for  Recurrence Time in Iterated Open Quantum Systems}
\author{P. Sinkovicz, T. Kiss, J. K. Asb\'oth}
\affiliation{Institute for Solid State Physics and Optics, 
Wigner Research Centre, Hungarian Academy of Sciences, 
H-1525 Budapest P.O. Box 49, Hungary}


\begin{abstract}
We consider recurrence to the initial state after repeated actions of
a quantum channel.  After each iteration a projective measurement is
applied to check recurrence. The corresponding return time is known to
be an integer for the special case of unital channels, including
unitary channels. We prove that for a more general class of quantum
channels the expected return time can be given as the inverse of the
weight of the initial state in the steady state. This statement is a
generalization of the Kac's lemma for classical Markov chains.
\end{abstract}
\maketitle


\section{\label{sec:Intro}Introduction}
According to the Poincar\'e recurrence
theorem\cite{poincare1890probleme, barreira2006poincare}, any closed
classical physical system, when observed for sufficiently long time,
will return arbitrarily close to its initial state.  The extension of
this statement to open systems whose dynamics can be modeled as a
Markov chains\cite{aldous2002reversible, kemeny1960finite}, is given
by Kac's lemma\cite{Kac47}. This connects the expected return time to
the limit distribution, i.e., the probability distribution of the
system in the infinite-time limit. The expected return time
(Poincar\'e time) is the inverse of the weight of the initial state in
the limit distribution. If the weight vanishes, the expected return
time diverges, and the probability of return is less than 1.

A version of the Poincar\'e recurrence theorem also holds for closed
quantum mechanical systems, which are periodically measured to test
whether recurrence has happened. Here the system is assumed to start
from an initial state given by a wavefunction, which is evolved by the
same unitary timestep operator between successive measurements. Every
measurement influences the dynamics: it either signals a return, and
projects the system to the initial state, or signals no return, and
projects out the initial state from the wavefunction.  Gr\"unbaum
\emph{et al.}\cite{Grunbaum13} have shown that in this case the
expected return time (number of measurements) is either infinite, or
it is an integer.  We generalized\cite{our14} this result to open
quantum mechanical systems, whose dynamics is given by a quantum
channel\cite{breuer2002theory, chruscinski2014time, alicki2001quantum,
  benatti2005open, paulsen2002completely}. We found that if the
channel is unital, i.e., if the limit distribution is the completely
mixed state in an effective Hilbert space that contains the initial
state\cite{arias2002fixed, holbrook2003noiseless,
  novotny2010asymptotic}, then the expected return time is equal to
the dimensionality of that effective Hilbert space.

One cannot help but notice the fact that the expected return time in
open quantum systems with unital dynamics is an integer is in
accordance with Kac's lemma. Indeed, the dimensionality of the
effective Hilbert space is the inverse of the weight of the initial
state in the (appropriately defined) limit distribution of the
dynamics. The question arises: can we push this statement further, is
there a broader class of open quantum dynamical system for which a
quantum Kac's lemma would hold?  In this paper we answer this question
positively. We show that for a rather general type of quantum channels
the expected return time can be calculated from the related steady
state of the system.

The structure of the paper is the following. Section \ref{sec:Defs}
contains the definitions and ideas that we need to set up the problem.
We state our result in Sect. \ref{sec:Theorem}.  In
Sect. \ref{example} we give a possible application of our theorem.
Section \ref{sec:Disc} concludes the paper with the discussion of the
results and some potentially interesting questions.


\section{Definitions\label{sec:Defs}}
We consider discrete-time dynamics of an open quantum system.  The
state of the system is described by a density operator $\hrho(t)$,
representing a mixture of pure states from an $N$-dimensional Hilbert
space $\mathcal{H}$, with $t\in\mathbb{N}$ denoting the discrete
time. Starting from a pure initial state $\ket{\rinitial} \in
\mathcal{H}$, we obtain the state by iterations of the fixed timestep
superoperator $\sss [\cdot]$,
\begin{equation}
\hrho(t) = \sss^t[\ket{\rinitial}\bra{\rinitial}]. 
\label{eq:dyn_def}
\end{equation}
As usual, we assume $\SSS{\cdot}$ to be a linear, trace-preserving and
completely positive map, i.e., a quantum channel. It can thus be
written using Kraus operators\cite{kraus1983states,
  sudarshan1961stochastic, choi2000completely, bengtsson2006geometry,
  wolf2012quantum} as
\begin{equation}
\SSS{\hrho} = \sum_{\nu = 0}^r \hA_\nu \hrho \hA_\nu^\dagger,
\label{eq:original_dyn}
\end{equation}
where $r \le N^2$ is the Kraus rank of $\sss[\cdot]$, and $\hA_\nu :
\mathcal{H} \to \mathcal{H}$ are the Kraus operators, with the
normalization
\begin{equation}
\sum_\nu \hA_\nu^\dagger \hA_\nu = \hI \, ,
\end{equation}
where $\hI$ represents the unit operator on $\mathcal{H}$.

We can construct a steady state of the dynamics from the initial state
$\ket{\rinitial}$ as
\begin{equation}
\hchi = \limes{T} \frac{1}{T} \sum_{t=0}^{T-1} \SSSn{\ket{\rinitial}\bra{\rinitial}} 
\, .
\label{eq:chi_Def}
\end{equation}
This limit is always well defined since we are in a finite dimensional
Hilbert space\cite{wolf2012quantum}.  The operator $\hchi$ is a convex
mixture of density operators, and is thus self-adjoint, positive, and
fulfils $\tr [\hchi] = 1$: it can be interpreted as a density operator
of the system. It represents a steady state, since
\begin{equation}
\SSS{\hchi} = \hchi - \limes{T} \frac{1}{T} \ket{\Psi}\bra{\Psi} = \hchi \, .
\label{eq:chi_fixedpoint}
\end{equation}


\subsection*{First return time}
To measure the first return time, we need to disturb the
dynamics\cite{kempe2003quantum,davies1976quantum}. We follow each
timestep by a dichotomic measurement that checks whether the system
has returned to the state $\ket{\rinitial}\bra{\rinitial}$, or if it
is in the orthogonal subspace $\mathcal{H}_{\perp}$, defined by
\begin{align}
\ket{\Phi} \in \mathcal{H}_\perp \quad \Leftrightarrow \quad
\bra{\Phi}\Psi\rangle = 0.
\label{eq:Hperp_def} 
\end{align}
The post-measurement state corresponding to ``no return'' is obtained
using the projector
\begin{equation}
\MMM{\hrho} = \left( \hI - 
\ket{\rinitial} \bra{\rinitial} \right)
\hrho \left(
\hI - \ket{\rinitial} \bra{\rinitial} \right) \, .
\end{equation}

Note that the projector $\mmm[\cdot]$ does not conserve the trace: its
outcome is a \emph{conditional density operator}, whose trace
represents the probability that the particle described by $\hrho$ has
not returned.  The modified dynamics, including the dichotomic
measurements, is defined by
\begin{equation}
\hrho_{\text{cond}}(t) = \MSt{\ket{\rinitial}\bra{\rinitial}} \, .
\label{cond_dyn_def}
\end{equation}
Here, the trace of the conditional density operator is a probability,
\begin{equation}
\tr \hrho_{\text{cond}}(t) = \text{P}(\text{no return up until time $t$}).
\label{cond_density_def}
\end{equation}

We call the iterated open quantum channel \emph{recurrent} when
started from $\ket{\Psi}$, if it returns with probability 1 in the
sense defined above, i.e., if $\lim \tr \hrho_\text{cond}(t)=0$. This
is in line with Ref.~\onlinecite{Grunbaum13}. In the rest of this
paper we will only concern ourselves with recurrent channels.

Let us remark that the states of a system evolving from the initial
state $\ket{\Psi}$, either by the evolution operator $\SSS{\cdot}$ or
by $\mmm\sss[\cdot]$, may explore only a subspace of $\mathcal{H}$,
but they span the same subspace, i.e., the relevant Hilbert space
$\mathcal{H}_{\text{rel}}$, as we have shown previously\cite{our14}.

The expected return time $\expectT$ is the expectation value of the
first return time, calculated using the probabilities in
Eq.~\eqref{cond_density_def}.  Whenever the iterated open quantum
channel is recurrent when started from $\ket{\Psi}$, the expected
return time can be expressed using the sum of the conditional density
operators\cite{our14}, which we denote by
\begin{equation}
\rcond = \limes{T} \sum_{t=0}^T \rcond(t) \,  .
\label{eq:def_rcond}
\end{equation}
Note that unlike in Eq.~\eqref{eq:chi_Def} for the steady state, there
is no factor of $1/T$ here, and so this sum can diverge.  If the sum
is divergent, than the expected return time is also divergent,
otherwise the trace of $\rcond$ gives the expected return time,
\begin{equation}
\expectT = \Tr \rcond \, .
\label{eq:expT_rcond}
\end{equation}

In Ref.~\onlinecite{our14} we have shown that in the case of unital
dynamics, where all eigenvalues of $\hchi$ are equal, the expected
return time $\expectT$ can be obtained from the steady state
$\hchi$. In that case, we found that the expected return time is an
integer, equal to the dimensionality of $\hchi$. In the next Section
we generalize this result, and give a formula that connects the
expected return time $\expectT$ to the steady state $\hchi$ for a
broader class of dynamical systems.


\section{Quantum Kac lemma\label{sec:Theorem}}
We now formulate the main result of this paper. If the initial state
$\ket{\rinitial}$ is an eigenvector of the steady state $\hchi$, with
nonzero eigenvalue $\lambda \in \mathbb{R}$, then the expected return
time is $\expectT = 1/\lambda$.  In formulas,
\begin{align}
\label{eq:expect_theorem}
\hchi \ket{\rinitial}&=\lambda \ket{\rinitial}, \quad \lambda \neq 0 &
\Longrightarrow \quad
\expectT &= \frac{1}{\bra{\rinitial} \hchi \ket{\rinitial}} \,  .
\end{align}
This is a direct generalization of the classical Kac's
lemma\cite{Kac47}, where the expected first return time to site $n$ is
the reciprocal of the corresponding component $\pi_n$ of the
equilibrium distribution vector $\pi$. We therefore refer to
Eq.~\eqref{eq:expect_theorem} as the quantum Kac lemma.

Note that $\lambda \neq 0$ ensures that $\lim \rcond(t) = 0$, proved
in Appendix \ref{app:recurrence}.


\subsection*{Proof\label{subsec:Proof}}

The quantum Kac lemma is a direct consequence of the statement,
\begin{equation}
\hchi \ket{\rinitial} = \lambda \ket{\rinitial} \,  \Leftrightarrow  \, \rcond = \frac{1}{\bra{\rinitial}\hchi\ket{\rinitial}} \hchi \, .
\label{eq:theorem}
\end{equation}
This states that the sum $\hrho_\text{cond}$ of the conditional
density operators, Eq.~(\ref{eq:def_rcond}), proportional to the
steady state $\hchi$ of Eq.~\eqref{eq:chi_Def}, if and only if
$\ket{\rinitial}$ is an eigenstate of $\hchi$.  The trace of the
relation on the right-hand-side of Eq.~\eqref{eq:theorem}, using
Eq.~\eqref{eq:expT_rcond}, gives directly the quantum Kac lemma,
Eq.~\eqref{eq:expect_theorem}. It remains to show that
Eq.~\eqref{eq:theorem} holds.

We now prove Eq.~\eqref{eq:theorem} in two steps.  

First, we show that whenever $\rcond$ is proportional to $\hchi$, then
the initial state is one of the eigenvectors of $\hchi$. This
statement follows from the fact that we can express $\rcond$ as
\begin{equation}
\rcond = \ket{\rinitial} \bra{\rinitial} + \limes{T} \sum_{t=1}^T \MSt{ \ket{\rinitial} \bra{\rinitial}} \, .
\label{eq:rcond_form}
\end{equation}
The conditional dynamics maps any density operator $\hrho$ to the
orthogonal subspace, that is $\MS{\hrho}: \mathcal{H} \to
\mathcal{H}_{\perp}$.  Furthermore, $\rcond$ is proportional to the
steady state, hence by applying Eq.~(\ref{eq:rcond_form}) to the
initial state $\ket{\rinitial}$ we can establish the following,
\begin{equation}
\lambda =\bra{\rinitial} \hchi \ket{\rinitial} \, .
\label{eq:eigenvalue}
\end{equation}
This concludes the first step of the proof.

The second step in the proof of Eq.~\eqref{eq:theorem} is to show,
that whenever the initial state $\ket{\Psi}$ is an eigenstate of the
density operator of the steady state $\hchi$, then $\rcond$ is
proportional to the steady state.  We notice, that if
$\ket{\rinitial}$ is an eigenvector of $\hchi = \hchi^\dagger$, with
eigenvalue $\lambda \neq 0$, then the steady state is given by
\begin{equation}
\hchi = \lambda \, \ket{\rinitial} \bra{\rinitial} +\hchi_\perp \, ,
\label{eq:chi_form}
\end{equation}
where $\hchi_\perp = \mmm[\hchi]$.  On the other hand, $\hchi$ is a
steady state of the dynamics, see Eq.~(\ref{eq:chi_fixedpoint}), i.e.,
$\hchi =\sss[\hchi]$. Projecting both sides of this latter equation
using $\mmm[\cdot]$, inserting Eq.~(\ref{eq:chi_form}), and
rearranging gives us
\begin{align}
\mathcal{MS} \big[ \ket{\Psi}\bra{\Psi} \big] &= \frac{1}{\lambda}
\left( \hchi_\perp - \MS{\hchi_\perp} \right) \, .
\label{eq:stilde}
\end{align}
Finally, inserting Eq.~\eqref{eq:stilde} in Eq.~(\ref{eq:rcond_form})
gives us 
\begin{align}
\rcond &= \ket{\rinitial} \bra{\rinitial} + \frac{1}{\lambda}
\lim_{T\to\infty} \sum_{t=0}^{T} \MSt{ \hchi_\perp - \MS{\hchi_\perp} } =
\nonumber\\ &= \frac{1}{\lambda} \hchi -
\frac{1}{\lambda}
\lim_{T\to\infty} \MSt{\hchi_\perp}.
\label{eq:almost_ready}
\end{align}

To finish the proof of the second step, we need to show that 
\begin{align}
\lim_{t\to\infty} \MSt{\hchi_\perp} = 0 \, . 
\label{eq:almost_ready}
\end{align}
This will be enough because by assumption, $\lambda \neq 0$. It is
clear that $\hchi_\perp$ is a unnormalized density operator in the
relevant Hilbert space. In fact, using the results of
Ref.~\onlinecite{liesen2012krylov}, it can be shown that the relevant
Hilbert space is already spanned by the first $N$ states in the orbit,
i.e., by the states $\rcond(n)$, with $n =
0,\ldots, N-1$, where $N$ is the dimensionality of the relevant
Hilbert space. In formulas,
\begin{align}
\hchi_\perp = \sum_{n=0}^{N-1} c_n (\mmm\sss)^n[\ket{\Psi}\bra{\Psi}] \, ,
\end{align}
with some complex coefficients $c_n\in \mathbb{C}$. Now consider
applying $(\mmm\sss)^t[\cdot]$ to this equation, and take the limit
$t\to\infty$. As we show in Appendix \ref{app:recurrence}, each term
on the right-hand-side vanishes. Therefore, the finite sum also
vanishes, and so we have Eq.~\eqref{eq:almost_ready}. This completes
the proof of Eq.~\eqref{eq:theorem}, and thus, of the quantum Kac
lemma.


\section{Example: evaluating hitting time via classical monitoring\label{example}}
Besides the class of unital dynamics, the quantum Kac lemma also
applies to any system where during each timestep the initial state
$\ket{\rinitial}$ is interfaced to the rest of the system only by
incoherent processes. More specifically, it applies when the timestep
operator $\sss[\cdot]$ can be split into three parts,
\begin{equation}
\label{eq:example_01}
\hrho(t+1) = \sss[\hrho(t)] = \ddd_\text{out} \left[ 
\ttt[ \ddd_\text{in} [\hrho(t)]]\right] \,. 
\end{equation}
Here, $\ttt[\cdot]$ is a superoperator that acts nontrivially only in
$\mathcal{H}_\perp$, defined in Eq.~\eqref{eq:Hperp_def}, i.e.,
\begin{equation}
\label{TTT_def}
\ttt[\hrho] = \ket{\Psi}\bra{\Psi}\hrho \ket{\Psi}\bra{\Psi}+ 
\sum_{\nu} \hK_\nu^{\phantom{\dagger}} \hrho
\hK_\nu^\dagger \, ,
\end{equation}
with Kraus operators $\hK_\nu:\mathcal{H} \to
\mathcal{H}_{\perp}$. The superoperators $\ddd_\text{in}[\cdot]$ and
$\ddd_\text{out}[\cdot]$ describe the incoherent particle transfer
from $\ket{\Psi}$ to the rest of the system and vice versa,
\begin{align}
\label{DDDinout}
\ddd_\text{in}[\hrho] &= 
\sum_{\nu=1}^N p_\nu \ket{\Phi_\nu}\bra{\Psi} \hrho 
\ket{\Psi}\bra{\Phi_\nu} + \hA_\text{in} \hrho \hA_\text{in} \, ;\\
\hA_\text{in} &=
\sqrt{\hI-\sum_{\nu=1}^N p_\nu \ket{\Psi}\bra{\Psi}} \, ;
\\
\ddd_\text{out}[\hrho] &= 
\sum_{\mu=1}^M q_\mu \ket{\Psi}\bra{\alpha_\mu} \hrho 
\ket{\alpha_\mu} \bra{\Psi} +
\hA_\text{out} \hrho \hA_\text{out} \, ;\\
\hA_\text{out} &=
\sqrt{\hI - \sum_{\mu=1}^M q_\mu \ket{\alpha_\mu}\bra{\alpha_\mu}} \, .
\end{align}
Here the rates $p_\nu,q_\mu$ are assumed to be positive, and $\sum
{p_\nu}\le 1$, and $\sum {q_\mu} \le 1$, and the states
$\ket{\Phi_\nu},\ket{\alpha_\mu}\in\mathcal{H}_\perp$.  In this case,
the condition in Eq.(\ref{eq:expect_theorem}) is automatically
satisfied, therefore we can apply the quantum Kac lemma and determine
$\expectT$ as the inverse of the weight of the initial state
$\ket{\Psi}$ in the steady state $\hchi$.

The example of the previous paragraph can be used to express the
\emph{hitting time} for an iterated quantum dynamical system in terms
of a stationary distribution. For this, we let $\mathcal{H}_\perp$
denote the Hilbert space where this quantum dynamics take place, and
$\ket{\Phi_1}$ and $\ket{\alpha_1}$ denote the pure states from which
and into which the hitting time is sought. We extend the Hilbert space
by the extra ancilla state $\ket{\Psi}$, set $N=1$, $p_1=1$, and
$M=1$, $q_1=1$.  The hitting time from $\ket{\Phi_1}$ to
$\ket{\alpha_1}$ is then given by the expected return time to
$\ket{\Psi}$.


\section{Discussion\label{sec:Disc}}
In this paper we found a relationship between the return time to an
initial pure state and the steady state for an iterated quantum
channel.  If the initial state is an eigenvector of that steady state,
then the reciprocal of the corresponding eigenvalue gives us the
expected return time for that particular initial state. This is not
only a generalization of the results of Ref.~\onlinecite{our14} for
unital channels (which includes unitary dynamics), but also of Kac's
lemma about Markov chains.

The condition $\hchi \ket{\rinitial} = \lambda \ket{\rinitial}$ is
sufficient, but not necessary, for the form of the expected return
time on the right-hand-side of Eq.~\eqref{eq:expect_theorem} to hold.
An example is given by the so-called classical-quantum channels
\cite{Shirokov13}, defined by the evolution equation
\begin{equation}
\hrho_{\text{cq}}(t+1)=\sss_{\text{cq}} \left[ \hrho(t) \right] = \sum_{n = 1}^{\text{dim}\mathcal{H}} \bra{\varphi_n} \hrho(t) \ket{\varphi_n} \, \hsigma_n \, ,
\end{equation}
where the generators of the dynamics $\{ \hsigma_n \}$ are positive
and self-adjoint operators with unit trace, and the Hilbert-space
vectors $\{\ket{\varphi_n} \}$ are an orthonormal set (we note that a
similar, but not equivalent notion of a classical-quantum channel was
introduced in \cite{lardizabal2015class}). The speciality of this type
of dynamics can be understood if we restrict our attention only to the
dynamics of the diagonal elements. The matrix elements within the
diagonal are transformed among themselves by a time independent
transition matrix $\hat{W}$, where $W_{m,n} = \bra{\varphi_m}
\hsigma_n \ket{\varphi_m}$ gives the probability that the system makes
the $\ket{\varphi_n} \to \ket{\varphi_m}$ transition. Based on
$\hat{W}$ one can naturally construct a classical homogeneous Markov
chain for which the original Kac's lemma is valid and has the same
recurrence properties.  One can generalize the previous example, and
say that Eq.~(\ref{eq:expect_theorem}) holds for every random walk,
where the evolution of the diagonal elements depend only on other
diagonal elements. In these cases the dynamics of the diagonal
elements can be separated from the dynamics of the off-diagonal
elements, and their evolution can be described as a discrete time
classical random walk, for which the classical Kac's lemma can be
applied.

In our generalization of Kac's lemma as well as in the example of the
classical-quantum channel the steady state corresponding to the given
initial state fully determines the expected return time. It would be
fascinating to know, whether there are some even more general classes
of quantum channels for which the knowledge of the steady state is
enough to calculate the first return time. 

Let us note that there are alternative ways to define a return time.
One can avoid the disturbance of the measurement, for example, by
taking a new system from an ensemble after each measurement. The
P\'olya number for quantum walks characterizing recurrence has been
defined in this way \cite{vstefavnak2008recurrence,
  vstefavnak2008recurrence02}.  There are also alternative ways to
define iterative open quantum dynamics, e.g., the ``open quantum
random walks''\cite{attal2012open}, for which there are known results
on recurrence and return time\cite{lardizabal2015class}.

Our theorem proved to be a useful tool to determine the hitting time
for an arbitrary iterated quantum channel by applying an extra
monitoring site coupled classically to the initial and final states of
the system to be observed. Monitoring the hitting time in this way is
a discrete-time analog for the hitting time defined for
continuous-time quantum walks, which defined through the survival time
of an excitation in the system where the Hamiltonian includes a
trapping site \cite{mulken2007survival, mulken2007quantum,
  varbanov2008hitting}.


\section{Acknowledgements}
We thank Gernot Alber, Andr\'as Frigyik and Melinda Herb\'ath for
stimulating discussions. We acknowledge support by the Hungarian
Scientific Research Fund (OTKA) under Contract Nos. K83858, NN109651,
the Hungarian Academy of Sciences (Lend\"ulet Program, LP2011-016) and
the Deutscher Akademischer Austauschdienst (Tempus-DAAD project
no. 65049).  J.K.A. was supported by the Janos Bolyai Scholarship of
the Hungarian Academy of Sciences.


\appendix
\section{Recurrence of the initial state\label{app:recurrence}}

In this section we prove that if the initial state $\ket{\Psi}$ is one
of the eigenvectors of the steady state $\hchi$ of
Eq.~\eqref{eq:chi_Def} with eigenvalue $\lambda \neq 0$, then $\lim \rcond(t) = 0$, in other words the process
is recurrent (returns with probability 1), i.e., 
\begin{align}
\label{eq:to_prove}
\hchi \ket{\Psi} &= \lambda \ket{\Psi}, \quad \lambda \neq 0
\quad \Longrightarrow \quad
\lim_{t\to\infty} (\mmm\sss)^t[\ket{\Psi}\bra{\Psi}] = 0.
\end{align}

For the proof, we use a steady state of the operator $\mmm\sss[\cdot]$
defined as
\begin{align}
\chim = \lim_{T\to\infty} \frac{1}{T} \sum_{t=0}^{T-1} 
(\mmm\sss)^t[\ket{\Psi}\bra{\Psi}]. 
\label{eq:def_chim}
\end{align}
The operator $\chim$ is a nonnegative, Hermitian operator, and it is
nonvanishing if $\lim \text{\Tr} \left\{
(\mmm\sss)^t[\ket{\Psi}\bra{\Psi}] \right\} > C \in \mathbb{R}^+$.  The
operator $\chim$ is obviously an unnormalized steady state of
$\mmm\sss[\cdot]$, i.e.,
\begin{align}
\mmm\sss[\chim] &= \chim. 
\end{align}  
Taking the expectation value of the two sides of this relation in the
state $\ket{\Psi}$ tells us 
\begin{align}
\label{eq:expect_chim_0}
\bra\Psi \chim \ket\Psi &= 0.
\end{align}  
On the other hand, $\bra\Psi \sss[\chim] \ket\Psi=0$ must hold as
well, or else the trace of $\chim$ would decrease under the mapping
$\mmm\sss[\cdot]$. Therefore, $\chim$ is not only a steady state of
$\mmm\sss[\cdot]$, but also of $\sss[\cdot]$, i.e.,
\begin{align}
\sss[\chim] &= \chim.
\end{align}  
We remark that the operator $\mmm[\cdot]$ does not take us out of the
relevant Hilbert space\cite{our14}, and thus the operator $\chim$ has
all its support in the relevant Hilbert space. 

We will prove Eq.~\eqref{eq:to_prove} below, by showing that 
$\Tr \chim = 0$. 

\subsubsection{Theoretical tools: Decaying subspace, minimal enclosures}

In the proof, we will use the concepts of minimal enclosures, and of
the decaying subspace, as introduced in
Ref.~\onlinecite{baumgartner2012structures}.  We recapitulate the
definitions, and the basic properties, below.

The decaying subspace $\mathcal{D}$ is defined as
\begin{equation}
\mathcal{D} = \{ \ket{\varphi} \in \mathcal{H}: \quad \forall \hrho \quad \lim_{t \to \infty} \bra{\varphi} \mathcal{S}^t[\hrho] \ket{\varphi}=0 \} \, ,
\end{equation}
i.e., it is spanned by the states which have a vanishing overlap in
the long time limit with any density operator $\hrho$ which acts in
$\mathcal{H}$. We will use $\mathcal{R}$ to denote its complement,
i.e.,
\begin{equation}
\mathcal{R} = \{ \ket{\varphi} \in \mathcal{H}: 
\quad \forall \ket{\phiD}\in
\mathcal{D} : \braket{\varphi}{\phiD}=0 \}.
\end{equation}

Time evolution by $\sss[\cdot]$ does not lead out of the set
$\mathcal{R}$, i.e., $\forall t\in\mathbb{N}, \ket{\phiR} \in
\mathcal{R}, \ket{\phiD} \in \mathcal{D}$:
\begin{equation}
\bra{\phiD}\, \sss^t[\ket{\phiR}\bra{\phiR}] \,\ket{\phiD} = 0.
\end{equation}
This property of $\mathcal{R}$ defines it to be an
enclosure\cite{baumgartner2012structures}. Like every enclosure, the
space $\mathcal{R}$ can be written as the sum of orthogonal minimal
enclosures\cite{baumgartner2012structures},
\begin{align}
\mathcal{R} &= \mathcal{R}_1 \oplus \mathcal{R}_2 \oplus \ldots \oplus 
\mathcal{R}_{M}.  
\end{align}

\subsubsection{The proof}

We begin by showing that the initial state $\ket{\Psi}$ lies in the
subspace $\mathcal{R}$, i.e.,
\begin{align}
\ket{\Psi}\in\mathcal{R}. 
\label{eq:first_step_app}
\end{align}
We show this by splitting the initial state into components from the
two subspaces,
\begin{equation}
\ket{\Psi} = \cD \ket{\PsiD} + \cR \ket{\PsiR} \, ,
\end{equation}
where $\cD,\cR\in\mathbb{C}$, and $\ket{\PsiD} \in \mathcal{D}$, and
$\ket{\PsiR} \in \mathcal{R}$. Using this decomposition, we have
\begin{align}
\bra{\PsiD}\hchi\ket{\Psi} = \cD \bra{\PsiD} \hchi \ket{\PsiD}
+ \cR \bra{\PsiD} \hchi \ket{\PsiR},
\end{align}
where $\hchi$ is the steady state of $\sss[\cdot]$ defined by
Eq.~\eqref{eq:chi_Def}.  The first term on the right-hand-side
vanishes, since $\hchi$ is a fixed point of $\sss[\cdot]$, and so it
can have no weight in the decaying subspace, We can bound the second
term from above using the Schwarz inequality for the vectors
$\hchi^{1/2}\ket{\PsiD}$ and $\hchi^{1/2}\ket{\PsiR}$, whereby
\begin{align}
\abs{\bra{\PsiD}\hchi\ket{\PsiR}}^2 \le 
\bra{\PsiD}\hchi\ket{\PsiD}
\bra{\PsiR}\hchi\ket{\PsiR} = 0,
\end{align}
where we used the positivity of $\hchi$ as well as the fact that
$\bra{\PsiD}\hchi\ket{\PsiD} = 0$.  Thus, $\bra{\PsiD}\hchi\ket{\Psi}
= 0$.  On the other hand, since $\hchi\ket{\Psi} = \lambda \ket\Psi$,
we have
\begin{align}
\bra{\PsiD}\hchi\ket{\Psi} &= 
\lambda \cD \braket{\PsiD}{\PsiD} + \lambda \cR \braket{\PsiD}{\PsiR} 
= \lambda \cD,
\end{align}
where we used the orthogonality of $\mathcal{D}$ and $\mathcal{R}$, as
well as the normalization of the vectors $\ket{\PsiD}$ and
$\ket{\PsiR}$.  Comparing this last result with
$\bra{\PsiD}\hchi\ket{\Psi} = 0$ gives us $\cD=0$, or, equivalently,
$\ket{\Psi}\in\mathcal{R}$, which is Eq.~\eqref{eq:first_step_app},
the first step.

Since $\ket{\Psi}\in\mathcal{R}$, the relevant Hilbert space is the
sum of a subset of the minimal enclosures, those that are not
orthogonal to the initial state $\ket{\Psi}$.  We let these be the
first $M'$ minimal orthogonal enclosures. We can then split the
initial states into components from these enclosures,
\begin{align}
\ket{\Psi} &= c_1\ket{\Psi_1} + c_2\ket{\Psi_2} + \ldots +
c_{M'}\ket{\Psi_{M'}},
\end{align}
with $c_j\in\mathbb{C}$, and $c_j\neq 0$ for all $j=1,\ldots,M'$.  The
unique steady states of $\sss[\cdot]$ in each of these minimal
enclosures is $\hchi_j$. Note that $\bra\Psi \hchi_j \ket\Psi >0$ for
all $j$. Any steady state of $\sss[\cdot]$ in the relevant Hilbert
space is a convex combination of the $\hchi_j$
\cite{baumgartner2012structures}. Since $\chim$ is an unnormalized
steady state of $\sss[\cdot]$ in the relevant Hilbert space, it can be
written as $\chim = \sum p_j \hchi_j$ with nonnegative weights $p_j\ge
0$. Therefore,
\begin{align}
\bra{\Psi}\chim\ket{\Psi} = \sum_{j=1}^{M'} 
p_j \bra{\Psi} \hchi_j \ket{\Psi}. 
\end{align}
Since $\bra\Psi \hchi_j \ket\Psi >0$ for all $j$, we must have $p_j=0$
for all $j$, i.e., the operator $\chim$ of Eq.~\eqref{eq:def_chim}
must vanish. This proves Eq.~(\ref{eq:to_prove}).


\bibliography{qkac}
\end{document}